\documentclass[showpacs,preprintnumbers,amsmath,amssymb,twocolumn]{revtex4}

\usepackage{graphicx}
\usepackage{dcolumn}
\usepackage{bm}

\begin{document}

\title{Relativistic BCS-BEC Crossover at Zero Temperature}
\author{\normalsize{Lianyi He and Pengfei Zhuang}}
\affiliation{Physics Department, Tsinghua University, Beijing
100084, China}

\begin{abstract}
We investigate the BCS-BEC crossover at zero temperature in the
frame of a relativistic model. The universality of the BCS-BEC
crossover for non-relativistic systems breaks down in relativistic
case and the crossover can be induced by changing the density.
When the effective scattering length is much less than the fermion
Compton wavelength, we recover the non-relativistic result if the
gas is initially in non-relativistic state. At ultra-strong
coupling where the scattering length is of the order of the
Compton wavelength, a new BEC state appears. In this state the
condensed bosons become nearly massless and anti-fermions are
excited. The behavior of the Goldstone mode and the mixing between
the amplitude and phase modes are significantly different in
different condensed regions.
\end{abstract}

\date{\today}

\maketitle

\section{Introduction}
\label{s1} It was found many years ago that the
Bardeen-Cooper-Shriffer (BCS) superfluidity/superconductivity in
degenerate Fermi gas and the Bose-Einstein condensation (BEC) of
composite molecules can be smoothly connected when Eagles and
Leggett found that the ground state wave functions of BCS and BEC
states are essentially the same\cite{BEC1,BEC2}. The BCS-BEC
crossover phenomena can be realized in high temperature
superconductivity and atomic Fermi gas where the s-wave scattering
length can be adjusted\cite{BEC3,BEC4,BEC5,BEC6}.

Recently, the study on BCS-BEC crossover has been extended to
relativistic Fermi systems with the Nozieres-Schmitt-Rink (NSR)
theory\cite{RBEC1,RBEC2} and the Bose-Fermi model\cite{RBEC3}. In
the NSR theory above the critical temperature, a new feature of
BCS-BEC crossover is found: There exist two kinds of BEC
states\cite{RBEC1}, the non-relativistic BEC (NBEC) and the
relativistic BEC (RBEC). In the RBEC state, the condensed bosons
become nearly massless, anti-matter can be excited, and matter and
anti-matter are nearly degenerate. A natural question is: Does the
RBEC appear only in relativistic systems and never exist in
non-relativistic systems? In this paper, we will argue that the
RBEC phase can happen in any system if the attractive coupling is
large enough, even though the system is initially in
non-relativistic state. From text books\cite{kerson}, when the
Fermi momentum of a system is much smaller than the fermion mass,
we can treat the system non-relativistically. This statement is
true only if the coupling is weak enough. For sufficiently strong
coupling, the system will become relativistic even though the
Fermi momentum is much smaller than the fermion mass.

We will extend in this paper the theory of BCS-BEC crossover in
symmetry breaking state\cite{BEC5} at zero temperature to
relativistic systems. The requirement for such an extension is
that we should recover the non-relativistic theory in a proper
limit. We will focus on two relativistic effects: the anti-fermion
degrees of freedom and the nontrivial fermion mass. The first
effect leads to the appearance of the RBEC state at sufficiently
strong coupling, and the second effect breaks the universality of
BCS-BEC crossover. We will also investigate the collective modes
evolution in the BCS-BEC crossover.

\section{Relativistic Theory of BCS-BEC Crossover}
\label{s2} The physical motivation why we need a relativistic
theory of BCS-BEC crossover is mostly due to the study of QCD
phase diagram, especially the dense quark matter which may exist
in compact stars and can be created in heavy ion collisions.
However, we will point out that the theory is also necessary for
non-relativistic systems when the coupling is strong enough. To
this end, let us first review the non-relativistic theory of
BCS-BEC crossover in a dilute Fermi gas.

Leggett's mean field theory\cite{BEC1} is successful to describe
the non-relativistic BCS-BEC crossover at zero temperature. For a
dilute Fermi gas with fixed density $n=k_f^3/(3\pi^2)$, where
$k_f$ is the Fermi momentum, the BCS-BEC crossover can be found if
one self-consistently solves the gap and number equations for the
pairing gap $\Delta_0$ and the fermion chemical potential $\mu_n$,
\begin{eqnarray}
-\frac{m}{4\pi a_s}&=&\int{d^3{\bf k}\over
(2\pi)^3}\left({\frac{1}{
2E_{\bf k}}-\frac{m}{{\bf k}^2}}\right),\nonumber\\
\frac{k_f^3}{3\pi^2}&=&\int{d^3{\bf k}\over
(2\pi)^3}\left(1-\frac{\xi_{\bf k}}{E_{\bf k}}\right),
\end{eqnarray}
where $E_{\bf k}$ and $\xi_{\bf k}$ are defined as $E_{\bf
k}=\sqrt{\xi_{\bf k}^2+\Delta_0^2}$ and $\xi_{\bf k}={\bf
k}^2/(2m)-\mu_n$, and $a_s$ is the s-wave scattering length. The
fermion mass $m$ plays a trivial role here, since the BCS-BEC
crossover depends only on the dimensionless parameter
$\eta=1/(k_fa_s)$, which is the so-called universality for
non-relativistic syetems. The BCS-BEC crossover can be
characterized by the chemical potential $\mu_n$: It coincides with
the Fermi energy $\epsilon_f=k_f^2/(2m)$ in the BCS limit
$\eta\rightarrow -\infty$ but becomes negative in the BEC region.
In the BEC limit $\eta\rightarrow+\infty$, one has
$\mu_n\rightarrow-E_b/2$ with $E_b=1/(ma_s^2)=2\eta^2\epsilon_f$
being the molecular binding energy. Therefore, in the
non-relativistic theory the chemical potential will tend to be
negatively infinity in the BEC limit.

A problem arises if we look into the physics of the BEC limit from
a relativistic point of view. In the relativistic description, the
fermion dispersion becomes $\xi_{\bf k}^\pm=\sqrt{{\bf
k}^2+m^2}\pm\mu_r$, where $\mp$ correspond to fermion and
anti-fermion degrees of freedom, and $\mu_r$ is the chemical
potential in relativistic theory\cite{kapusta}. In the
non-relativistic limit $|{\bf k}|\ll m$, if $\mu_r$ is of the
order of $m$, we can neglect the anti-fermion degrees of freedom
and recover the non-relativistic dispersion $\xi_{\bf
k}^-\simeq{\bf k}^2/(2m)-(\mu_r-m)$. Therefore, not $\mu_r$ itself
but the quantity $\mu_r-m$ plays the role of non-relativistic
chemical potential $\mu_n$\cite{kapusta}. While the chemical
potential $\mu_n$ can be arbitrarily negative in non-relativistic
theory, $\mu_r$ is under some physical constraint in relativistic
theory. Since the molecule binding energy can not be larger than
two times the constituent mass, the absolute value of the
non-relativistic chemical potential $\mu_n=\mu_r-m=-E_b/2$ at
strong enough coupling can not exceed the fermion mass $m$, and
the relativistic chemical potential $\mu_r$ should be always
positive. If the number density $n$ satisfies $k_f\ll m$, the
non-relativistic theory works well when the coupling is not very
strong and the binding energy satisfies $E_b\ll 2m$. However, if
the coupling is strong enough to ensure $E_b\sim 2m$, relativistic
effects will appear. From $E_b\simeq1/(ma_s^2)$, we can roughly
estimate that the non-relativistic theory becomes unphysical for
the s-wave scattering length $a_s\sim1/m$. This can be understood
if we consider $1/m$ as the Compton wavelength $\lambda_c$ of a
particle.

What will then happen at $\eta\rightarrow+\infty$ in an attractive
Fermi gas?  At sufficiently strong coupling with $E_b\rightarrow
2m$ and $\mu_r\rightarrow 0$, the dispersions $\xi_{\bf k}^-$ and
$\xi_{\bf k}^+$ for fermions and anti-fermions become nearly
degenerate, and non-relativistic limit can not be reached even for
systems with $k_f\ll m$. This means that the anti-fermion pairs
can be excited and become nearly degenerate with the fermion
pairs, and the condensed bosons and anti-bosons become nearly
massless. Without any model dependent calculation, we can confirm
an important relativistic effect in BCS-BEC crossover: There
should exist a relativistic BEC(RBEC) state\cite{RBEC1} which is
smoothly connected to the non-relativistic BEC(NBEC) state. The
RBEC is not a specific phenomenon for relativistic fermion
systems, it should appear in any fermion system if the attractive
coupling becomes strong enough, even though the initial
non-interacting gas satisfies $k_f\ll m$.

We now start to construct a general relativistic model, which we
expect to recover the non-relativistic theory in a proper limit.
We consider only fermion degrees of freedom in the original
Lagrangian. The most general form of the Lagrangian density can be
expressed as
\begin{equation}
{\cal L}=\bar{\psi}\left(i\gamma^\mu\partial_\mu-m\right)\psi
+{\cal L}_I ,
\end{equation}
where $\psi,\bar{\psi}$ denote the Dirac fields with mass $m$, and
${\cal L}_I$ describes the attractive interaction among the
fermions. It is shown that the dominant interaction is in the
scalar channel $J^P=0^+$\cite{RBCS,RBCS2}. For the sake of
simplicity, we consider the zero range interaction and take the
form
\begin{equation}
{\cal L}_I=\frac{g}{4}\left(\bar{\psi} i\gamma_5C\bar{\psi}^{\text
T}\right)\left(\psi^{\text T}C i\gamma_5\psi\right) ,
\end{equation}
where $g$ is the attractive coupling constant, and
$C=i\gamma_0\gamma_2$ is the charge conjugation matrix. Generally,
by increasing the attractive coupling, the crossover from
condensation of spin-zero Cooper pairs at weak coupling to the
Bose-Einstein condensation of bound bosons at strong coupling can
be realized.

We start our calculation from the partition function in imaginary
time formalism,
\begin{equation}
Z=\int D\bar{\psi} D\psi e^{\int_0^\beta d\tau\int d^3{\bf
x}({\cal L}+\mu\psi^\dagger\psi)},
\end{equation}
where $\beta$ is the inverse temperature, $\beta=1/T$, and $\mu$
is the chemical potential corresponding to the net charge density
$\psi^\dagger\psi$ which is determined by the charge conservation.
Similar to the method in the study of superconductivity, we
introduce the Nambu-Gorkov spinors
\begin{equation}
\Psi = \left(\begin{array}{cc} \psi\\
C\bar{\psi}^{\text{T}}\end{array}\right),\ \ \ \bar{\Psi} =
\left(\begin{array}{cc} \bar{\psi} &
\psi^{\text{T}}C\end{array}\right)
\end{equation}
and the auxiliary pair field $\Delta(x)=g\psi^{\text T}(x)C
i\gamma_5\psi(x)/2$, and then perform the Hubbard-Stratonovich
transformation, the partition function can be written as
\begin{equation}
Z=\int D\bar{\Psi} D\Psi D\Delta D\Delta^* e^{\int_0^\beta
d\tau\int d^3{\bf x}\left(\frac{1}{2}\bar{\Psi}{\bf
G}^{-1}\Psi-\frac{|\Delta|^2}{g}\right)}
\end{equation}
in terms of the inverse Nambu-Gorkov propagator
\begin{eqnarray}
\label{g} {\bf
G}^{-1}=i\gamma^\mu\partial_\mu-m+\mu\gamma_0\sigma_3+i\gamma_5\Delta\sigma_++i\gamma_5\Delta^*\sigma_-,
\end{eqnarray}
where $\sigma_\pm=(\sigma_1\pm i\sigma_2)/2$ are defined in
Nambu-Gorkov space with $\sigma_i(i=1,2,3)$ being the Pauli
matrices. Integrating out the fermion fields, we finally derive
the partition function
\begin{equation}
Z=\int D\Delta D\Delta^*\ e^{-S_{eff}}
\end{equation}
with the bosonic effective action
\begin{eqnarray}
\label{eff} S_{eff}=\int_0^\beta d\tau\int d^3{\bf
x}\left[\frac{|\Delta|^2}{g}-\frac{1}{2\beta}\text{Tr}\ln
[\beta{\bf G}^{-1}]\right].
\end{eqnarray}

While in general case one should include the contribution from
pair fluctuations\cite{BEC3,BEC4,BEC6}, the Leggett mean field
theory is already good to describe the BCS-BEC crossover at zero
temperature. This can be seen from the fact that, the dominant
contribution of fluctuations to the thermodynamic potential at low
temperature is from the massless Goldstone modes and is
proportional to $T^4$. At zero temperature it vanishes\cite{BEC5}.
While the mean field theory can describe well the BCS-BEC
crossover, it can not predict precise values of the universal
constants in the unitary limit and the boson-boson scattering
length in the BEC limit. Since our goal in this paper is to
investigate the BCS-BEC crossover only, we will take the mean
field approximation in the following.

In the mean field approximation, we consider the uniform and
static saddle point $\Delta(x)=\Delta_0$ which satisfies the
stationary condition $\delta S_{eff}[\Delta_0]/\delta \Delta_0=0$.
The thermodynamic potential density
$\Omega=S_{eff}[\Delta_0]/(\beta V)$ at the saddle point can be
evaluated as
\begin{eqnarray}
\Omega ={\Delta_0^2\over g}-\int{d^3 {\bf k}\over
(2\pi)^3}\left(E_{\bf k}^-+E_{\bf k}^+-\xi_{\bf k}^--\xi_{\bf
k}^+\right)
\end{eqnarray}
where we have defined the notations $E_{\bf k}^\pm=\sqrt{(\xi_{\bf
k}^\pm)^2+\Delta_0^2}$ and $ \xi_{\bf k}^\pm=\epsilon_{\bf
k}\pm\mu$ with $\epsilon_{\bf k}=\sqrt{{\bf k}^2+m^2}$. Minimizing
$\Omega$, we get the gap equation for the condensate $\Delta_0$,
\begin{equation}
\label{gap} \frac{1}{g}={1\over 2}\int{d^3{\bf k}\over
(2\pi)^3}\left(\frac{1}{E_{\bf k}^-}+\frac{1}{E_{\bf k}^+}\right)
\end{equation}
and the number equation for the fermion density
$n=k_f^3/(3\pi^2)$,
\begin{equation}
\label{number} \frac{k_f^3}{3\pi^2}=\int{d^3{\bf k}\over
(2\pi)^3}\left[\left(1-\frac{\xi_{\bf k}^-}{E_{\bf
k}^-}\right)-\left(1-\frac{\xi_{\bf k}^+}{E_{\bf
k}^+}\right)\right].
\end{equation}
The first and second terms on the right hand side of Equations
(\ref{gap}) and (\ref{number}) correspond to fermion and
anti-fermion degrees of freedom, respectively.

The model is non-renormalizable and a proper regularization is
needed. We subtract the vacuum contribution
$\partial\Omega/\partial\Delta_0^2|_{T=n=\Delta_0=0}$ from the gap
equation, namely we replace the bare coupling $g$ by a
renormalized coupling $U$,
\begin{equation}
-\frac{1}{U}=\frac{1}{g}-\frac{1}{2}\int{d^3{\bf k}\over
(2\pi)^3}\left(\frac{1}{\epsilon_{\bf k}-m}+\frac{1}{\epsilon_{\bf
k}+m}\right).
\end{equation}
To recover the non-relativistic theory correctly, the chemical
potential in vacuum should be chosen as $\mu=m$ rather than
$\mu=0$. Such a subtraction is consistent with the formula derived
from the two body scattering matrix\cite{RBEC2}. The effective
s-wave scattering length $a_s$ can be defined as $U=4\pi a_s/m$.
While this is a natural extension of the regularization in
non-relativistic theory, the ultraviolet divergence can not be
completely removed, and a momentum cutoff $\Lambda$ still exists
in the theory. The relativistic Fermi energy $E_f$ in the theory
can be define as $E_f=\sqrt{k_f^2+m^2}$, which recovers the Fermi
kinetic energy $E_f-m\simeq\epsilon_f= k_f^2/(2m)$ in
non-relativistic limit.

\section {Relativistic Effects in BCS-BEC Crossover}
\label{s3} In the non-relativistic theory, there are only two
characteristic lengths $k_f^{-1}$ and $a_s$, and the BCS-BEC
crossover shows the universality: After a proper scaling, all
physical quantities depend only on the dimensionless coupling
$\eta=1/(k_f a_s)$. Especially, in the unitary limit
$a_s\rightarrow\infty$, all scaled physical quantities become
universal constants. Unlike the non-relativistic theory where the
fermion mass $m$ plays a trivial role, in the relativistic theory
a new length scale, namely the Compton wavelength
$\lambda_c=m^{-1}$ appears. As a consequence, the BCS-BEC
crossover should depend on not only the dimensionless coupling
$\eta$, but also the quantity $\zeta=k_f/m=k_f\lambda_c$. Since
the cutoff $\Lambda$ is needed, there is also a $\Lambda/m$
dependence. By scaling all energies by $\epsilon_f$ and momenta by
$k_f$, the gap and number equations (\ref{gap}) and (\ref{number})
become dimensionless,
\begin{eqnarray}\label{scale}
-\frac{\pi}{2}\eta &=& \int_0^zx^2dx\bigg[\left(\frac{1}{E_x^-}-\frac{1}{\epsilon_x-2\zeta^{-2}}\right)\nonumber\\
&&\ \ \ \ \ \ \ \ \ \ +\left(\frac{1}{E_x^+}-\frac{1}{\epsilon_x+2\zeta^{-2}}\right)\bigg],\nonumber\\
\frac{2}{3}&=&\int_0^zx^2dx\bigg[\left(1-\frac{\xi_x^-}{E_x^-}\right)-\left(1-\frac{\xi_x^+}{E_x^+}\right)\bigg]
\end{eqnarray}
with the definitions
$E_x^\pm=\sqrt{\left(\xi_x^\pm\right)^2+\left(\Delta_0/\epsilon_f\right)^2}$,
$\xi_x^\pm=\epsilon_x\pm\mu/\epsilon_f$,
$\epsilon_x=2\zeta^{-1}\sqrt{x^2+\zeta^{-2}}$ and
$z=\zeta^{-1}\Lambda/m$. It becomes now clear that the BCS-BEC
crossover is characterized by three dimensionless parameters,
$\eta, \zeta$ and $\Lambda/m$.

We now study when we can recover the non-relativistic theory in
the limit $\zeta\ll1$. Expanding $\epsilon_x$ in powers of
$\zeta$, $\epsilon_x=x^2+2\zeta^{-2}+O(\zeta^2)$, we can recover
the non-relativistic version of $\xi_x$. However, we can not
simply neglect the terms corresponding to anti-fermions, namely
the second terms on the right hand side of Equations
(\ref{scale}). Such terms can be neglected only when $|\mu-m|$ and
$\Delta_0$ are both much less than $m$. When the coupling is so
strong to ensure $\mu\rightarrow 0$, the contribution from
anti-fermions becomes significant. In this case the condition
$\zeta\ll 1$ is not sufficient to describe the non-relativistic
limit, and the other important condition should be $a_s\ll 1/m$
which guarantees the molecule binding energy $E_b\ll 2m$. With the
dimensionless coupling $\eta$, we have $\eta=1/(k_f
a_s)=(m/k_f)(1/m a_s)\ll m/k_f =\zeta^{-1}$. Therefore, the
complete condition for the non-relativistic limit can be expressed
as $\zeta\ll 1$ and $\zeta\eta\ll 1$.

\begin{figure}[!htb]
\begin{center}
\includegraphics[width=7cm]{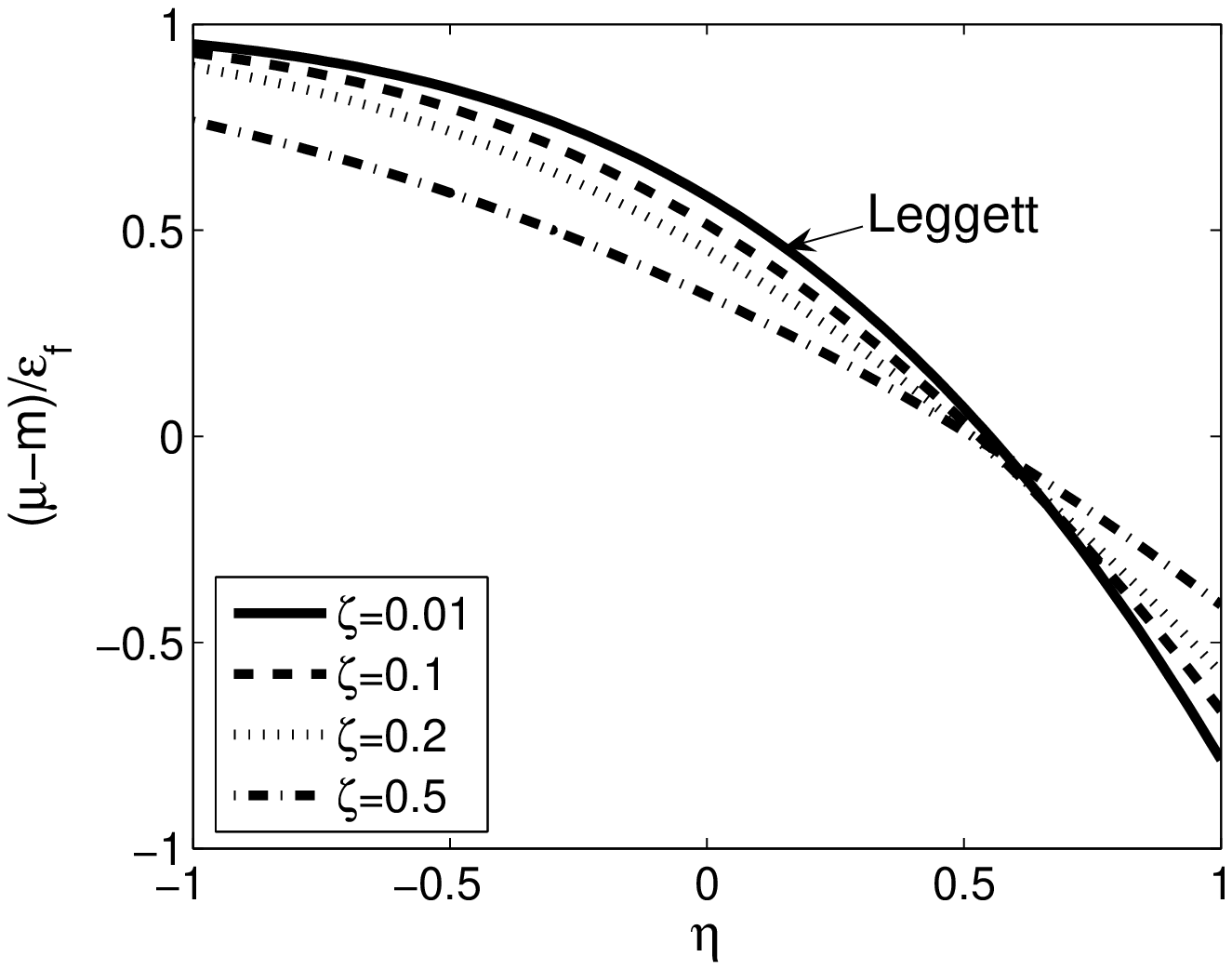}
\includegraphics[width=7cm]{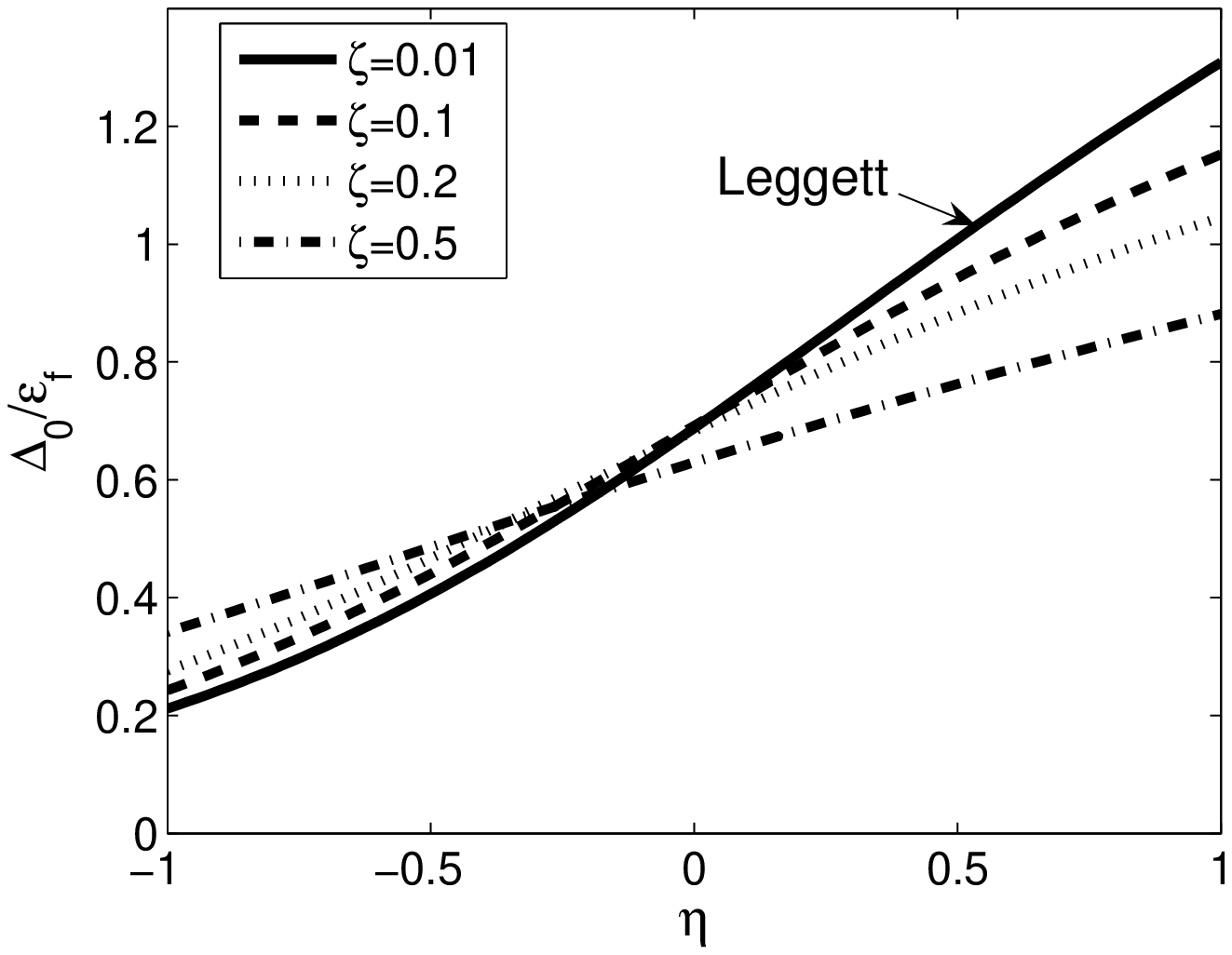}
\caption{The condensate $\Delta_0$ and non-relativistic chemical
potential $\mu-m$, scaled by the non-relativistic Fermi energy
$\epsilon_f$, as functions of $\eta$ in the region $-1<\eta<1$ for
several values of $\zeta$. In the calculations we set
$\Lambda/m=10$. \label{fig1}}
\end{center}
\end{figure}

\begin{figure}[!htb]
\begin{center}
\includegraphics[width=7cm]{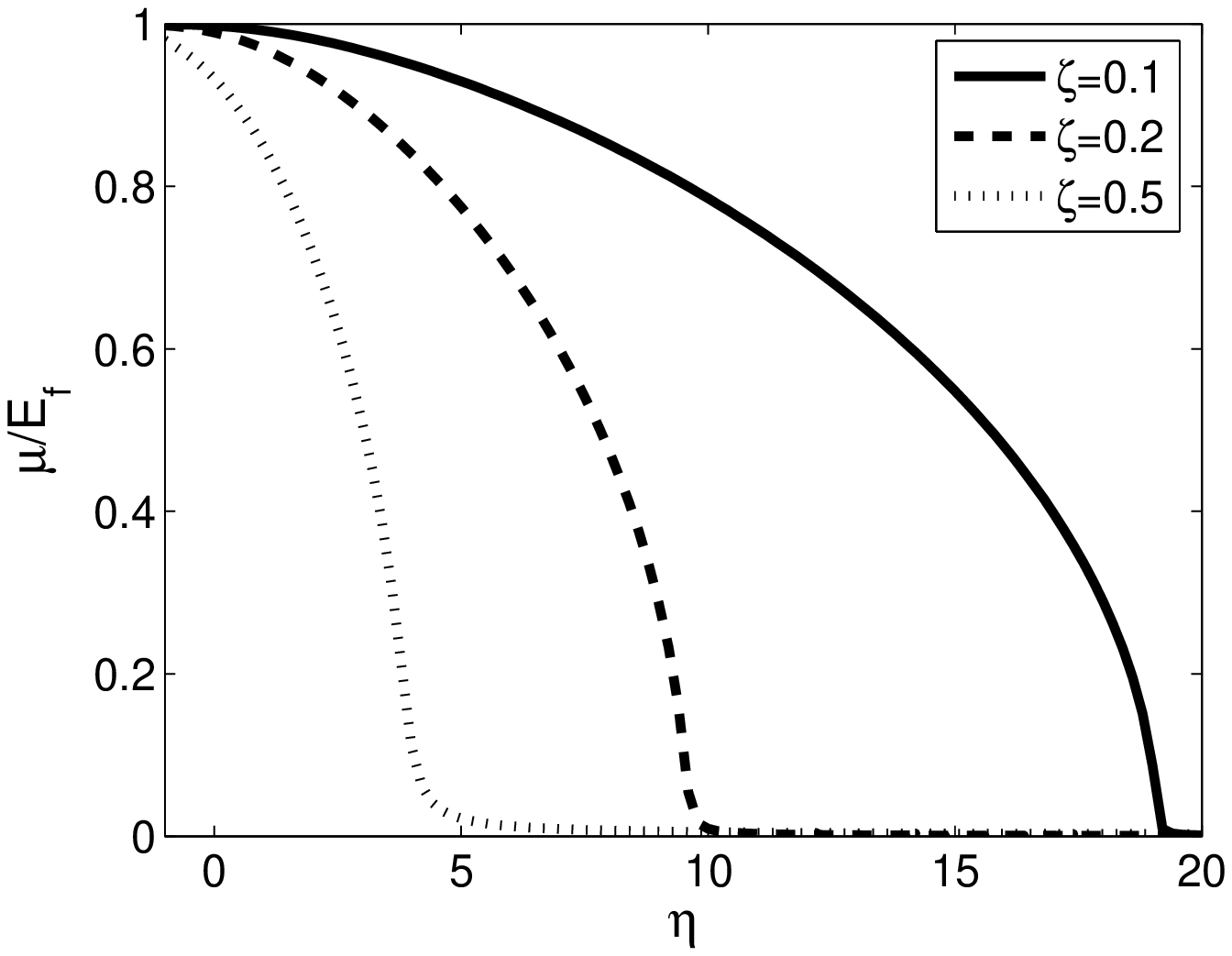}
\includegraphics[width=7cm]{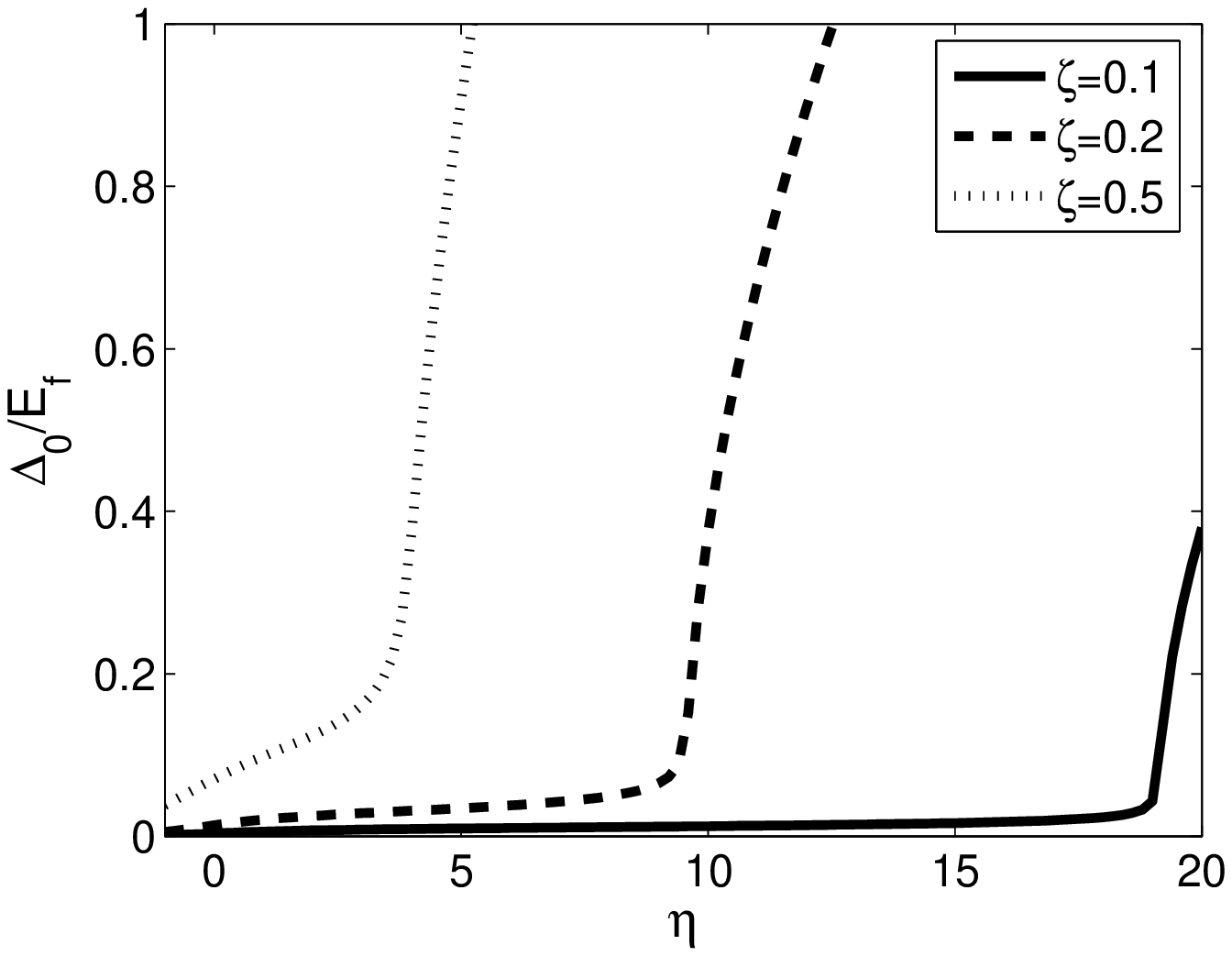}
\caption{The condensate $\Delta_0$ and chemical potential $\mu$,
scaled by the relativistic Fermi energy $E_f$, as functions of
$\eta$ in a wide region $-1<\eta<20$ for several values of
$\zeta$. In the calculations we set $\Lambda/m=10$. \label{fig2}}
\end{center}
\end{figure}

To confirm the above statement we solve the gap and number
equations numerically. In Fig.\ref{fig1} we show the condensate
$\Delta_0$ and non-relativistic chemical potential $\mu-m$ as
functions of $\eta$ in the region $-1<\eta<1$ for several values
of $\zeta<1$. In this region the cutoff dependence is weak and can
be neglected. For sufficiently small $\zeta$, we really recover
the Leggett result. With increasing $\zeta$, however, the
universality is broken and the deviation becomes more and more
remarkable. On the other hand, when we increase the coupling
$\eta$, especially for $\eta \rightarrow\zeta^{-1}$, the
difference between our calculation at any fixed $\zeta$ and the
Leggett result becomes significant due to relativistic effects.
This means that even in the case $\zeta\ll 1$ we can not recover
the non-relativistic result when the coupling $\eta$ is of the
order of $\zeta^{-1}$. This can be seen clearly from the $\eta$
dependence of $\Delta_0$ and $\mu$ in a wider $\eta$ region, shown
in Fig.\ref{fig2}. We found in our numerical calculation with the
cutoff $\Lambda/m=10$ a critical coupling $\eta_c\simeq
2\zeta^{-1}$ which is consistent with the above estimation. Beyond
this point the chemical potential $\mu$ approaches zero and the
condensate $\Delta_0$ becomes of the order of the relativistic
Fermi energy $E_f$ or the fermion mass $m$ rapidly. In the region
around $\eta_c$, the relativistic effect becomes significant, even
though the initial non-interacting gas is in a non-relativistic
state.

We can derive an analytical expression for the critical coupling
$\eta_c$ or $U_c$ for the RBEC state. At the critical coupling, we
can take $\mu\simeq 0$ and $\Delta_0\ll m$ approximately, and the
gap equation becomes
\begin{equation}
\frac{1}{U_c}\simeq\frac{m^2}{2\pi^2}\int_0^\Lambda
\frac{dk}{\sqrt{k^2+m^2}},
\end{equation}
from which we have
\begin{eqnarray}
U_c^{-1}&=&\frac{2}{\pi}U_0^{-1}f(\Lambda/m),\nonumber\\
\eta_c&=&\frac{2}{\pi}\zeta^{-1}f(\Lambda/m)
\end{eqnarray}
with the definitions $f(x)=\ln(x+\sqrt{x^2+1})$ and
$U_0=4\pi/m^2=4\pi\lambda_c^2$. While in the non-relativistic
region with $\eta\ll\zeta^{-1}$ the solution is almost cutoff
independent, in the RBEC region with $\eta\sim \zeta^{-1}$ the
solution becomes sensitive to the cutoff.

To see what happens in the region with $\eta\zeta\sim 1$, we
discuss the fermion and anti-fermion momentum distributions
$n_-({\bf k})$ and $n_+({\bf k})$,
\begin{equation}
n_\pm({\bf k})=\frac{1}{2}\left(1-\frac{\xi_{\bf k}^\pm}{E_{\bf
k}^\pm}\right).
\end{equation}
In the non-relativistic BCS and BEC regions with $\eta\zeta\ll1$,
we have $\Delta_0\ll E_f$, and the anti-fermion degrees of freedom
can be safely neglected, $n_+({\bf k})\simeq 0$. In the BCS limit
with $\Delta_0\ll \epsilon_f$, $n_-({\bf k})$ deviates slightly
from the standard Fermi distribution at the Fermi surface,
especially we have $n_-({\bf 0})\simeq 1$. In the deep BEC region,
we have $\Delta_0\sim\eta\epsilon_f$ and
$|\mu-m|\sim\eta^2\epsilon_f$ and in turn $|\mu-m|\gg\Delta_0$ and
$n_-({\bf 0})\ll 1$, and $n_-({\bf k})$ becomes very smooth.
However, in the RBEC region with $\eta\zeta\sim 1$, $\mu$
approaches zero, and the anti-fermions become nearly degenerate
with the fermions,
\begin{equation}
n_-({\bf k})\simeq n_+({\bf
k})=\frac{1}{2}\left(1-\frac{\epsilon_{\bf k}}{\sqrt{\epsilon_{\bf
k}^2+\Delta_0^2}}\right),
\end{equation}
where $n_\pm({\bf 0})$ can be large enough as long as $\Delta_0$
is of the order of $m$.

In the non-relativistic BCS and BEC regions, the total net density
$n=n_--n_+$ is approximately $n\simeq n_-=\int d^3{\bf k}/(2\pi)^3
n_-({\bf k})$, and the contribution from the anti-fermions can be
neglected, $n_+=\int d^3{\bf k}/(2\pi)^3 n_+({\bf k})\simeq 0$.
However, when we approach to the RBEC region, the contributions
from fermions and anti-fermions are almost equally important. In
the RBEC region with $\Delta_0<m$ we can estimate
\begin{eqnarray}
n_-\simeq n_+&\simeq&\frac{\Delta_0^2}{8\pi^2}\int_0^\Lambda
dk\frac{k^2}{k^2+m^2}\nonumber\\
&=&\frac{\Delta_0^2\Lambda}{8\pi^2}\left(1-\frac{m}{\Lambda}\arctan{\frac{\Lambda}{m}}\right).
\end{eqnarray}
For $m\ll\Lambda$, the second term in the bracket can be omitted,
$n_-\simeq n_+\simeq\Delta_0^2\Lambda/(8\pi^2)$. The densities of
fermions and anti-fermions are both much larger than $n$, and
their difference produces a conserved net density.

\section {Density Effect in BCS-BEC Crossover}
\label{s4} In the non-relativistic BCS-BEC crossover, the coupling
strength and number density are reflected in the theory in a
compact way through the dimensionless quantity $\eta=1/(k_f a_s)$,
and changing the density of the system can not induce a BCS-BEC
crossover. However, if the universality is broken, there would
exist an extra density dependence which may induce a BCS-BEC
crossover. In non-relativistic Fermi gas, the breaking of the
universality can be induced by a finite range
interaction\cite{density}. In the relativistic theory, the
universality is naturally broken by the $\zeta=k_f/m$ dependence
which leads to the extra density effect.

\begin{figure}[!htb]
\begin{center}
\includegraphics[width=7cm]{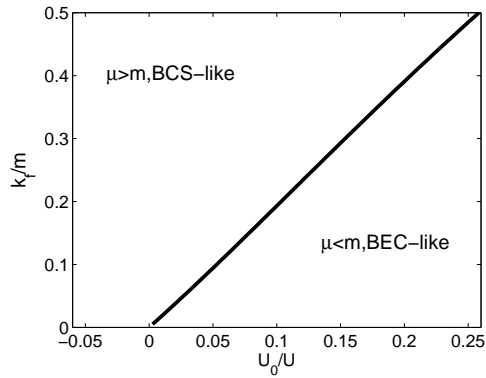}
\caption{The phase diagram in the $U_0/U-k_f/m$ plane. The line is
defined as $\mu=m$. \label{fig3}}
\end{center}
\end{figure}

To study this phenomenon, we calculate the phase diagram in the
$U_0/U-k_f/m$ plane where $U_0/U$ reflects the pure coupling
constant effect and $k_f/m$ reflects the pure density effect. The
reason why we do not present the phase diagram in the $\eta-\zeta$
plane is that both $\eta=1/(k_f a_s)$ and $\zeta=k_f/m$ include
the density effect. The phase diagram is shown in Fig.\ref{fig3}.
The BEC state is below the line $\mu=m$ and the BCS-like state is
above the line. We see clearly two ways to realize the BCS-BEC
crossover, by changing the coupling constant at a fixed density
and changing the density at a fixed coupling. Note that we only
plot the line which separates the two regions with $\mu>m$ and
$\mu<m$. Above and close to the line there should exist a
crossover region, like the phase diagram in \cite{density}.

The density induced BCS-BEC crossover can be realized in dense QCD
such as QCD at finite isospin density\cite{ISO1,ISO2,ISO3,ISO4}
and two color QCD at finite baryon density. The new feature in QCD
is that the quark mass $m$ decreases with increasing density due
to chiral symmetry restoration at finite density, which would
lower the line in the phase diagram Fig.\ref{fig3}.

\section{Collective Mode Evolution}
\label{s5}
To investigate the Gaussian fluctuations above the
saddle point $\Delta_0$, we write $\Delta(x)=\Delta_0+\phi(x)$ and
expand the action to the second order in $\phi$ to obtain
\begin{equation}
S_{\text{Gauss}}[\phi]=S_{\text{eff}}[\Delta_0]+\frac{1}{2}\sum_Q\Phi^\dagger(Q)
{\bf M}(Q)\Phi(Q),
\end{equation}
where $\Phi$ is defined as $\Phi^\dagger(Q)=(\phi^*(Q),\phi(-Q))$,
$Q=({\bf q},i\nu_m)$ with $\nu_m=2m\pi/\beta\
(m=0,\pm1,\pm2,\cdots)$ is the four momentum of the collective
mode, and $\sum_Q=T\sum_m\int d^3{\bf q}/(2\pi)^3$ denotes
integration over the three momentum ${\bf q}$ and summation over
the frequency $\nu_m$. The inverse propagator ${\bf M}$ of the
collective mode is a $2\times2$ matrix with the elements
\begin{eqnarray}
M_{11}(Q)&=& \frac{1}{g}+\frac{1}{2}\sum_K\text{Tr}
\left[i\gamma_5{\cal G}_0^{11}(K+Q)i\gamma_5{\cal G}_0^{22}(K)\right],\nonumber\\
M_{12}(Q)&=& \frac{1}{2}\sum_K\text{Tr} \left[i\gamma_5{\cal
G}_0^{12}(K+Q)i\gamma_5{\cal
G}_0^{12}(K)\right],\nonumber\\
M_{21}(Q)&=&M_{12}(-Q),\nonumber\\
M_{22}(Q)&=&M_{11}(-Q),
\end{eqnarray}
where $K=({\bf k},i\omega_n)$ with $\omega_n=(2n+1)\pi/\beta\
(n=0,1,2,\cdots)$ is the fermion four momentum, and ${\cal
G}_0^{ij} (i,j=1,2)$ are the elements of the fermion propagator
${\cal G}_0={\bf G}[\Delta_0]$ in the Nambu-Gorkov space given by
\begin{eqnarray}
{\cal G}_0^{11}&=& {i\omega_n+\xi_{\bf k}^-\over
(i\omega_n)^2-(E_{\bf k}^-)^2}\Lambda_+\gamma_0+ {i\omega_n-\xi_{\bf k}^+\over (i\omega_n)^2-(E_{\bf k}^+)^2}\Lambda_-\gamma_0,\nonumber\\
{\cal G}_0^{12}&=& {-i\Delta_0\over (i\omega_n)^2-(E_{\bf
k}^-)^2}\Lambda_+\gamma_5+ {-i\Delta_0\over (i\omega_n)^2-(E_{\bf
k}^+)^2}\Lambda_-\gamma_5,\nonumber\\
{\cal G}_0^{22}&=&{\cal G}_0^{11}(\mu\rightarrow-\mu),\nonumber\\
{\cal G}_0^{21}&=&{\cal G}_0^{12}(\mu\rightarrow-\mu)
\end{eqnarray}
with the energy projectors
\begin{equation}
\Lambda_{\pm}({\bf k}) = {1\over
2}\left[1\pm{\gamma_0\left(\vec{\gamma}\cdot{\bf k}+m\right)\over
\epsilon_{\bf k}}\right].
\end{equation}
At zero temperature, $M_{11}$ and $M_{12}$ can be evaluated as
\begin{eqnarray}
M_{11}&=&\frac{1}{g}+\int{d^3{\bf k}\over
(2\pi)^3}\\
&\Bigg[&\left(\frac{(v_{\bf k}^-)^2(v_{{\bf k}+{\bf
q}}^-)^2}{i\nu_m-E_{\bf k}^--E_{{\bf k}+{\bf q}}^-}-\frac{(u_{\bf
k}^-)^2(u_{{\bf k}+{\bf q}}^-)^2}{i\nu_m+E_{\bf k}^-+E_{{\bf
k}+{\bf q}}^-}\right){\cal
T}_+\nonumber\\
&+&\left(\frac{(u_{\bf k}^+)^2(u_{{\bf k}+{\bf q}}^+)^2}{i\nu_m-E_{\bf k}^+-E_{{\bf k}+{\bf q}}^+}
-\frac{(v_{\bf k}^+)^2(v_{{\bf k}+{\bf q}}^+)^2}{i\nu_m+E_{\bf k}^++E_{{\bf k}+{\bf q}}^+}\right){\cal T}_+\nonumber\\
&+&\left(\frac{(v_{\bf k}^-)^2(u_{{\bf k}+{\bf
q}}^+)^2}{i\nu_m-E_{\bf k}^--E_{{\bf k}+{\bf q}}^+}-\frac{(u_{\bf
k}^-)^2(v_{{\bf k}+{\bf q}}^+)^2}{i\nu_m+E_{\bf k}^-+E_{{\bf
k}+{\bf q}}^+}\right){\cal
T}_-\nonumber\\
&+&\left(\frac{(u_{\bf k}^+)^2(v_{{\bf k}+{\bf
q}}^-)^2}{i\nu_m-E_{\bf k}^+-E_{{\bf k} +{\bf q}}^-}-\frac{(v_{\bf
k}^+)^2(u_{{\bf k}+{\bf q}}^-)^2}{i\nu_m+E_{\bf k}^++E_{{\bf k}
+{\bf q}}^-}\right){\cal T}_-\Bigg],\nonumber\\
M_{12}&=&\int{d^3{\bf k}\over (2\pi)^3}\nonumber\\
&\Bigg[&\left(\frac{u_{\bf k}^-v_{\bf k}^-u_{{\bf k}+{\bf
q}}^-v_{{\bf k}+{\bf q}}^-}{i\nu_m+E_{\bf k}^-+E_{{\bf k}+{\bf
q}}^-}-\frac{u_{\bf k}^-v_{\bf k}^-u_{{\bf k}+{\bf q}}^-v_{{\bf
k}+{\bf q}}^-}{i\nu_m-E_{\bf k}^--E_{{\bf k}+{\bf
q}}^-}\right){\cal
T}_+\nonumber\\
&+&\left(\frac{u_{\bf k}^+v_{\bf k}^+u_{{\bf k}+{\bf q}}^+v_{{\bf
k}+{\bf q}}^+}{i\nu_m+E_{\bf k}^++E_{{\bf k}+{\bf
q}}^+}-\frac{u_{\bf k}^+v_{\bf k}^+u_{{\bf k}+{\bf q}}^+v_{{\bf
k}+{\bf q}}^+}{i\nu_m-E_{\bf k}^+-E_{{\bf k}+{\bf
q}}^+}\right){\cal T}_+
\nonumber\\
&+&\left(\frac{u_{\bf k}^-v_{\bf k}^-u_{{\bf k}+{\bf q}}^+v_{{\bf
k}+{\bf q}}^+}{i\nu_m+E_{\bf k}^-+E_{{\bf k}+{\bf
q}}^+}-\frac{u_{\bf k}^-v_{\bf k}^-u_{{\bf k}+{\bf q}}^+v_{{\bf
k}+{\bf q}}^+}{i\nu_m-E_{\bf k}^--E_{{\bf k}+{\bf
q}}^+}\right){\cal
T}_-\nonumber\\
&+&\left(\frac{u_{\bf k}^+v_{\bf k}^+u_{{\bf k}+{\bf q}}^-v_{{\bf
k}+{\bf q}}^-}{i\nu_m+E_{\bf k}^++E_{{\bf k}+{\bf
q}}^-}-\frac{u_{\bf k}^+v_{\bf k}^+u_{{\bf k}+{\bf q}}^-v_{{\bf
k}+{\bf q}}^-}{i\nu_m-E_{\bf k}^+-E_{{\bf k}+{\bf
q}}^-}\right){\cal T}_-\Bigg]\nonumber
\end{eqnarray}
with ${\cal T}_\pm=1/2\pm({\bf k}\cdot{\bf q}+\epsilon_{\bf
k}^2)/(2\epsilon_{\bf k}\epsilon_{{\bf k}+{\bf q}})$ and the
coherent coefficients $(u_{\bf k}^{\pm})^2=(1+\xi_{\bf
k}^\pm/E_{\bf k}^\pm)/2$ and $(\ v_{\bf k}^{\pm})^2=(1-\xi_{\bf
k}^\pm/E_{\bf k}^\pm)/2$.

Taking the analytical continuation $i\nu_m\rightarrow\omega+i0^+$,
the dispersion $\omega=\omega({\bf q})$ of the collective mode is
obtained by solving the equation
\begin{equation}
\det{{\bf M}[{\bf q},\omega({\bf q})]}=0.
\end{equation}
To make the result more physical, we decompose the complex
fluctuation field $\phi(x)$ into its amplitude mode $\lambda(x)$
and phase mode $\theta(x)$,
$\phi(x)=(\lambda(x)+i\theta(x))/\sqrt{2}$. Then the fluctuation
part of the effective action takes the form
\begin{equation}
\left(\begin{array}{cc} \lambda^*&\theta^*\end{array}\right)\left(\begin{array}{cc} M_{11}^++M_{12}&iM_{11}^-\\
-iM_{11}^- &
M_{11}^+-M_{12}\end{array}\right)\left(\begin{array}{c} \lambda\\
\theta\end{array}\right),
\end{equation}
where $M_{11}^\pm({\bf q},\omega)=(M_{11}({\bf q},\omega)\pm
M_{11}({\bf q},-\omega))/2$ are even and odd functions of
$\omega$. From $M_{11}^-({\bf q},0)=0$, the amplitude and phase
modes decouple exacrtly at $\omega=0$. Furthermore,
$M_{11}^+(0,0)=M_{12}(0,0)$ ensures that the phase mode at ${\bf
q}=0$ is gapless, i.e., the Goldstone mode.

We now determine the velocity of the Goldstone mode, $\omega({\bf
q})=c|{\bf q}|$ for $\omega,|{\bf q}|\ll \text{min}_{\bf
k}\{E_{\bf k}^\pm\}$. For this purpose, we make a small ${\bf q}$
and $\omega$ expansion of the effective action,
\begin{eqnarray}
M_{11}^++M_{12} &=& A+C|{\bf q}|^2-D\omega^2+\cdots,\nonumber\\
M_{11}^+-M_{12} &=& Q|{\bf q}|^2-R\omega^2+\cdots,\nonumber\\
M_{11}^- &=& B\omega+\cdots.
\end{eqnarray}
The Goldstone mode velocity now reads
\begin{equation}
c^2={Q\over B^2/A+R},
\end{equation}
and the corresponding eigenvector of ${\bf M}$ is
$(\lambda,\theta)=(-ic|{\bf q}|B/A,1)$, which is a pure phase mode
at ${\bf q}=0$ but has an admixture of the amplitude mode
controlled by $B$ at finite ${\bf q}$. The explicit form of $A,\
B,\ R$ and $Q$ can be calculated as
\begin{eqnarray}
A &=& 4\Delta_0^2 R,\nonumber\\
B &=& {1\over 4}\int{d^3{\bf k}\over
(2\pi)^3}\left({\xi_{\bf k}^-\over E_{\bf k}^{-3}}-{\xi_{\bf k}^+\over E_{\bf k}^{+3}}\right),\nonumber\\
R &=& {1\over 8}\int{d^3{\bf k}\over
(2\pi)^3}\left({1\over E_{\bf k}^{-3}}+{1\over E_{\bf k}^{+3}}\right),\nonumber\\
Q &=&{1\over 16}\int{d^3{\bf k}\over
(2\pi)^3}\Bigg[\frac{1}{E_{\bf k}^{-3}}\left(\frac{\xi_{\bf
k}^-}{\epsilon_{\bf k}}+\left(\frac{\Delta_0^2}{E_{\bf k}^{-2}}
-\frac{\xi_{\bf k}^-}{3\epsilon_{\bf k}}\right)\frac{{\bf k}^2}{\epsilon_{\bf k}^2}\right)\nonumber\\
&+&\frac{1}{E_{\bf k}^{+3}}\left(\frac{\xi_{\bf
k}^+}{\epsilon_{\bf k}}+\left(\frac{\Delta_0^2}{E_{\bf
k}^{+2}}-\frac{\xi_{\bf k}^+}{3\epsilon_{\bf k}}\right)
\frac{{\bf k}^2}{\epsilon_{\bf k}^2}\right)\nonumber\\
&+&2\left(\frac{1}{E_{\bf k}^-}+\frac{1}{E_{\bf k}^+}-2\frac{E_{\bf k}^-E_{\bf k}^+-\xi_{\bf k}^-\xi_{\bf k}^++\Delta_0^2}{E_{\bf k}^-E_{\bf k}^+(E_{\bf k}^-+E_{\bf k}^+)}\right)\nonumber\\
&\times&\frac{1}{\epsilon_{\bf k}^2}\left(1-\frac{{\bf
k}^2}{3\epsilon_{\bf k}^2}\right)\Bigg].
\end{eqnarray}

In the non-relativistic limit with $\zeta\ll 1$ and
$\zeta\eta\ll1$, we have $|\mu-m|\ll m$, all the terms that
include anti-fermion energy can be neglected, the fermion
dispersions $E_{\bf k}^-$ and $\xi_{\bf k}^-$ can be approximated
by $E_{\bf k}$ and $\xi_{\bf k}$, and the function ${\cal T}_+$
can be approximated by $1$. In this limit the functions $M_{11}$
and $M_{12}$ are the same as the ones obtained in non-relativistic
theory\cite{BEC5}. In this case, we have
\begin{eqnarray}
B &=& {1\over 2}\int{d^3{\bf k}\over
(2\pi)^3}{\xi_{\bf k}\over E_{\bf k}^3},\nonumber\\
R &=& {1\over 8}\int{d^3{\bf k}\over (2\pi)^3}{1\over E_{\bf k}^3},\nonumber\\
Q &=& {1\over 16}\int{d^3{\bf k}\over (2\pi)^3}{1\over E_{\bf
k}^3}\left({\xi_{\bf k}\over m}+{\Delta_0^2\over E_{\bf k}^2}{{\bf
k}^2\over m^2}\right).
\end{eqnarray}
In the BCS limit, all the integrated functions peak near the Fermi
surface, we have $B=0$ and  $c^2=Q/R$. Working out the integrals
we recover the well known result $c=\zeta/\sqrt{3}$ for
non-relativistic BCS superfluidity. In the BEC region, the Fermi
surface does not exist and $B$ becomes non-zero. An explicit
calculation shows that
$c=\zeta/\sqrt{3\pi\eta}\ll\zeta$\cite{BEC5}. This result can be
rewritten as $c^2=4\pi n_Ba_B/m_B^2$ where $m_B=2m,\ a_B=2a_s$ and
$n_B=n/2$ are corresponding mass, scattering length and density of
a dilute Bose gas, which recovers the result of the Bogoliubov
theory of dilute Bose gas with short-range repulsive
interaction\cite{bog}.

In the RBEC region with $\zeta\eta\sim1$, we have $\mu\rightarrow
0$, and the terms include anti-fermion energy become nearly
degenerate with the fermion terms and hence can not be neglected.
Since the quantity $B$ vanishes at $\mu\rightarrow0$, we have
$c^2=\lim_{\Lambda\rightarrow\infty} Q/R$ in the RBEC region.
Taking $\mu=0$ which leads to
\begin{eqnarray}
R &=& {1\over 4}\int{d^3{\bf k}\over (2\pi)^3}{1\over
E_{\bf k}^3},\nonumber\\
Q &=& {1\over 8}\int{d^3{\bf k}\over (2\pi)^3}{1\over E_{\bf
k}^3}\left(3-{{\bf k}^2\over\epsilon_{\bf k}^2}+{\Delta_0^2\over
E_{\bf k}^2}{{\bf k}^2\over\epsilon_{\bf k}^2}\right),
\end{eqnarray}
where $E_{\bf k}=\sqrt{\epsilon_{\bf k}^2+\Delta_0^2}$ is now the
degenerate dispersions in the limit $\mu\rightarrow 0$, we find
$c=1$ in this region. This is consistent with the Goldstone boson
velocity calculated from the relativistic boson field theory of
BEC\cite{bose}.

In the ultra-relativistic BCS state with $k_f\gg m$ and
$\Delta_0\ll\mu\simeq E_f$, all the terms that include
anti-fermion energy can be neglected again. In this case we have
$B=0$ and
\begin{eqnarray}
R &=& {\mu^2\over 16\pi^2}\int_0^\infty dk {1\over
\left[(k-\mu)^2+\Delta_0^2\right]^{3/2}},\nonumber\\
Q &=& {\mu^2\over 32\pi^2}\int_0^\infty dk{\Delta_0^2\over
\left[(k-\mu)^2+\Delta_0^2\right]^{5/2}}.
\end{eqnarray}
A simple algebra shows that $Q/R=3$, and hence we obtain the
well-known result $c=1/\sqrt{3}$ for the ultra-relativistic BCS
superfluidity.

The mixing of amplitude and phase modes is quite different in
different regions. In the weak coupling BCS region, all the
integrated functions peak near the Fermi surface and we have $B=0$
due to the particle-hole symmetry, and hence the amplitude and
phase modes decouple exactly. In the NBEC phase with
$\zeta\eta\ll1$, while the anti-fermion term in $B$ can be
neglected, we have $B\neq 0$ since the particle-hole symmetry is
lost, and the mixing is strong. In the RBEC region, while both
particle-hole and anti-particle--anti-hole symmetries are lost,
they cancel each other and we have again $B=0$. This can be seen
from the fact that for $\mu\rightarrow 0$ the first and second
terms in $B$ cancel to each other. Thus in the RBEC region, the
amplitude and the phase modes decouple again. This can also be
explained in the frame of the boson field theory of BEC. In
non-relativistic theory, the off-diagonal elements of the inverse
propagator are proportional to $i\omega$\cite{nao}, which induces
a strong mixing between amplitude and phase modes. However, in
relativistic theory, the off-diagonal elements are proportional to
$i\mu\omega$\cite{kapusta,bose}, which vanishes when BEC of
massless bosons happens.

\section{Summary}
\label{s6} We have investigated the BCS-BEC crossover at zero
temperature in a relativistic Fermi gas model. Unlike the
non-relativistic theory, the fermion mass plays a nontrivial role
and serves as a new length scale. As a consequence, the
universality breaks down and there exists an extra number density
effect on the BCS-BEC crossover. When the effective scattering
length is much less than the fermion Compton wavelength and the
Fermi momentum is much less than the fermion mass, we can recover
the non-relativistic theory. At the ultra-strong coupling where
the effective scattering length is of the order of the Compton
wavelength, the RBEC state appears. In this state the condensed
boson becomes nearly massless and anti-fermions are excited. The
sound velocity of the Goldstone mode and the mixing between the
amplitude and phase modes are quite different in different
regions, and the results agree well with the boson field theory.

In this paper, we investigated only the relativistic BCS-BEC
crossover at zero temperature where all pairs get condensed. At
finite temperature, zero momentum condensed pairs can be thermally
excited and one should go beyond the mean field theory to treat
properly the non-condensed pairs\cite{BEC6}. The study on the
relativistic BCS-BEC crossover in the symmetry breaking phase at
finite temperature is in progress.

{\bf Acknowledgement:}  This work is supported by the grants
NSFC10428510, 10435080 10575058 and SRFDP20040003103.

\end{document}